\documentclass{tMOP2e}

\newcommand{\D}{\mathop{\mathrm{d}}}

\begin{document}
\title{Statistical properties of the radiation from SASE FEL operating in a
post-saturation regime with and without undulator tapering}

\author{E.A. Schneidmiller, M.V. Yurkov, DESY, Hamburg, Germany }

\maketitle

\begin{abstract}

We describe statistical and coherence properties of the radiation from x-ray
free electron lasers (XFEL) operating in the post-saturation regime. We
consider practical case of the SASE3 FEL at the European XFEL. We perform
comparison of the main characteristics of the X-ray FEL operating in the
post-saturation regime with and without undulator tapering: efficiency,
coherence time and degree of transverse coherence.

\end{abstract}

\section{Introduction}

Radiation from Self Amplified Spontaneous Emission Free Electron Laser (SASE
FEL) \cite{dks-sase,pellegrini-murphy-sase} has limited spatial and temporal
coherence. This happens due to start-up of the amplification process from the
shot noise in the electron beam. The fluctuations of the electron beam density
are uncorrelated in time and space, and many radiation modes are excited at the
initial stage of amplification. As a rule, ground spatial TEM$_{00}$ mode with
highest gain dominates, and the degree of transverse coherence grows in the
exponential amplification stage. Radiation wave slips forward with respect to
the electron beam by one wavelength per one undulator period. This relative
slippage on the scale of the field gain length gives an estimate for coherence
length. Both, the degree of transverse coherence and the coherence time reach
maximum value in the end of the exponential regime of amplification, and then
degrade visibly in the nonlinear regime. Maximum degree of transverse coherence
of about 0.95 is reached for the values of the diffraction parameter about 1.
For large values of the diffraction parameter the degree of transverse
coherence falls down due to poor mode selection, i.e. mode degeneration takes
place. For small values of the diffraction parameter the degree of transverse
coherence falls down due to a poor longitudinal coherence
\cite{trcoh-oc,coherence-oc,coherence-anal-oc,coherence-njp}.

Radiation from SASE FEL with planar undulator contains visible contribution of
the odd harmonics. Parameter range where intensity of higher harmonics is
defined mainly by the nonlinear beam bunching in the fundamental harmonic has
been intensively studied in refs.
\cite{hg-1,hg-2,hg-2a,hg-3,hg-4,hg-5,hg-6,kim-1,kim-2,harm-prst}. Comprehensive
studies of the nonlinear harmonic generation have been performed in
\cite{harm-prst} in the framework of the one-dimensional model. General
features of the harmonic radiation have been determined. It was found that the
coherence time at the saturation falls inversely proportional to the harmonic
number, and relative spectrum bandwidth remains constant with the harmonic
number. Comprehensive study of the coherence properties of the odd harmonics in
the framework of the three-dimensional model have been performed in
\cite{harm-stat-3d-fel2012}. We considered parameter range when the intensity
of higher harmonics is mainly defined by the nonlinear harmonics generation
mechanism. The case of the optimized XFEL has been considered. Using similarity
techniques we derived universal dependencies for the main characteristics of
the SASE FEL covering all practical range of optimized X-ray FELs.

Application of the undulator tapering \cite{kroll-tapering} allows to increase
the conversion efficiency to rather high values
\cite{paladin-tap,fawley-scharl,fawley-2lg,fawley-vinokur,ssy-1993,handbook-91,physrep-95,book,litho-kw-jm3,bnl-tapering}.
It is used now as a routine procedure at LCLS and
SACLA, and is planned for use at SWISS FEL, PAL XFEL, and European XFEL.
\cite{lcls,sacla,euro-xfel-tdr,swiss-fel,pal-xfel}. Simulation studies of
the tapered SASE FELs have been performed (see
\cite{desy11-152,wu-tap-2012,gel-tap-2011,agapov-2014} and references therein).

There are two main reasons why we performed the present study. The first reason
is that the SASE3 FEL at the European XFEL, operating at long wavelengths, can
not be tuned as optimized FEL \cite{coherence-oc} due to the limitation on the
minimum value of the focusing beta function. Another reason is that in the
parameter range of SASE3 FEL the linear mechanism of harmonic generation is
essential which results in much higher power of the higher odd harmonic with
respect to the case of the nonlinear harmonic generation
\cite{desy11-152,sy-harm}.In our study we compare coherence properties of the
fundamental and the third harmonic for the case of untapered undulator and the
undulator with optimized tapering \cite{opt-tap-fel2014}. We show that the
brilliance of the fundamental harmonic of the radiation from the SASE FEL with
the optimized undulator tapering can be increased by a factor of 3 with respect
to untapered case.

\section{General definitions and simulation procedure}

The first order time correlation function $g_1(t,t')$ and the first-order
transverse correlation function $\gamma_1 (\vec{r}_{\perp}, \vec{r}\prime
_{\perp}, z, t) $ are defined as

\begin{eqnarray}
&& g_1(\vec{r},t-t')  =
\frac{\langle \tilde{E}(\vec{r},t)\tilde{E}^*(\vec{r},t')\rangle }
{\left[\langle \mid\tilde{E}(\vec{r},t)\mid^2\rangle
\langle \mid\tilde{E}(\vec{r},t')\mid^2\rangle \right]^{1/2}} \ ,
\nonumber \\
&& \gamma_1 (\vec{r}_{\perp}, \vec{r}\prime _{\perp}, z, t) = \frac{
\langle \tilde{E} (\vec{r}_{\perp}, z, t)
\tilde{E}^{*} (\vec{r}\prime _{\perp}, z, t) \rangle }
{ \left[ \langle |\tilde{E} (\vec{r}_{\perp}, z, t) |^2 \rangle
\langle |\tilde{E} (\vec{r}\prime _{\perp}, z, t) |^2 \rangle \right]^{1/2}}
\ ,
\end{eqnarray}

\noindent where $\tilde{E}$ is the slowly varying amplitude of the amplified
wave. For a stationary random process the time correlation function is the
function of arguments $\tau = t - t \prime$, and the first-order transverse
correlation function $\gamma_1 $ does not depend on time. The coherence time
and the degree of transverse coherence are defined as:

\begin{eqnarray}
&&
\tau_{\mathrm{c}} = \int
\limits^{\infty}_{-\infty} | g_1(\tau) |^2 \D\tau \ .
\nonumber \\
&&
\zeta =
\frac
{
\int |\gamma _1 (\vec{r}_{\perp},\vec{r}\prime _{\perp})|^{2}
I(\vec{r}_{\perp})
I(\vec{r}\prime _{\perp})
\D\vec{r}_{\perp}
\D\vec{r}\prime _{\perp}
}
{
[\int I(\vec{r}_{\perp}) \D\vec{r}_{\perp}]^{2}
} \ ,
\label{eq:def-degcoh}
\end{eqnarray}

\noindent where $I(\vec{r}_{\perp}) = \langle |\tilde{E} (\vec{r}_{\perp}) |^2
\rangle $. Peak brilliance is defined as a transversely coherent spectral flux:

\begin{equation}
B =
\frac{\omega \D \dot{N}_{ph}}{\D \omega} \
\frac{\zeta}{\left(\lambda/2\right)^2} \ .
\end{equation}

\begin{figure}[tb]

\includegraphics[width=0.45\textwidth]{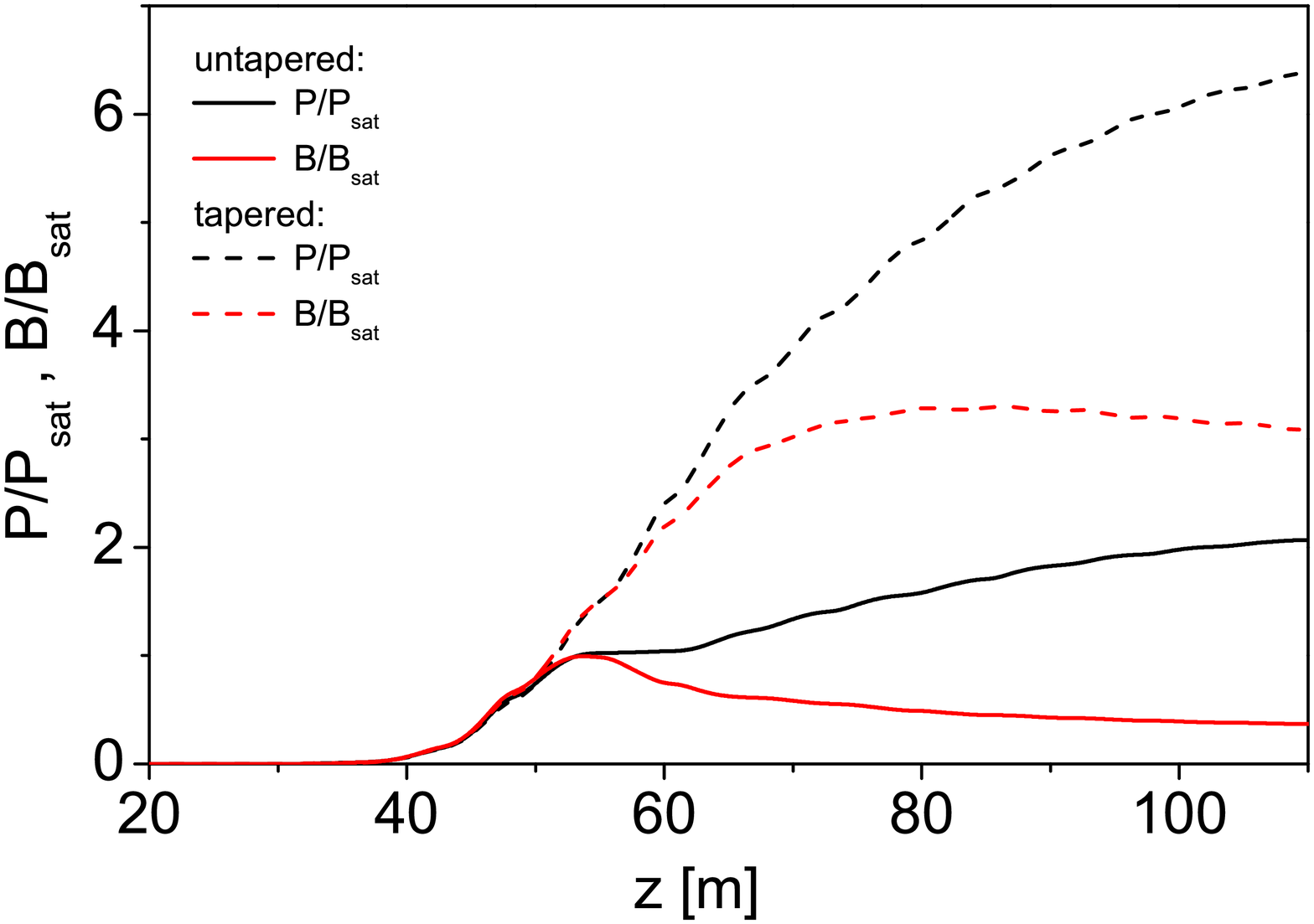}
\includegraphics[width=0.45\textwidth]{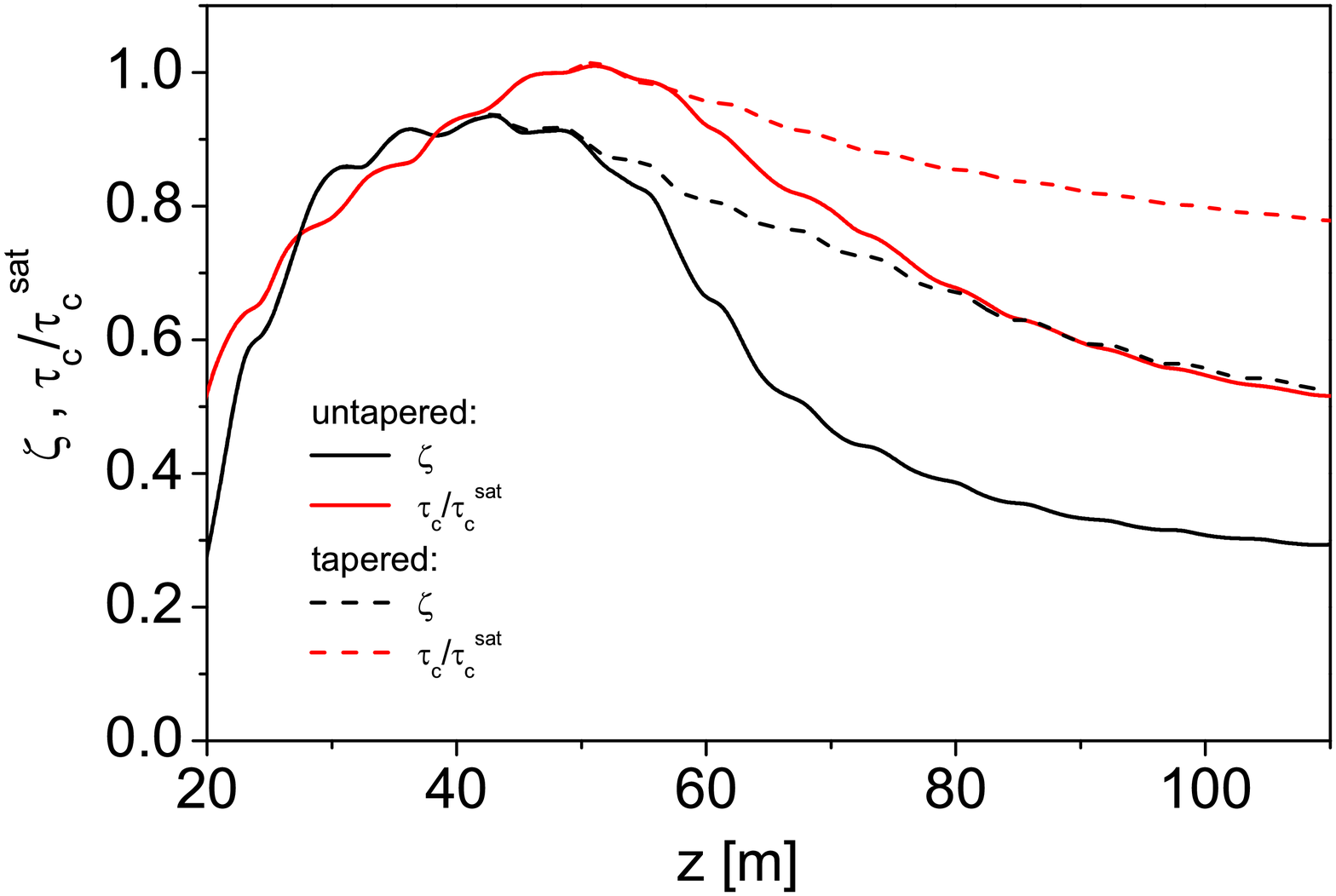}

\caption{
Fundamental harmonic:
evolution of the radiation power and brilliance (left plot) and of
coherence time and degree of transverse coherence (right plot) along the
undulator for untapered (solid curves) and optimized tapered case (dashed
curves). } \label{fig:pz1} \end{figure}

\begin{figure}[tb]

\includegraphics[width=0.45\textwidth]{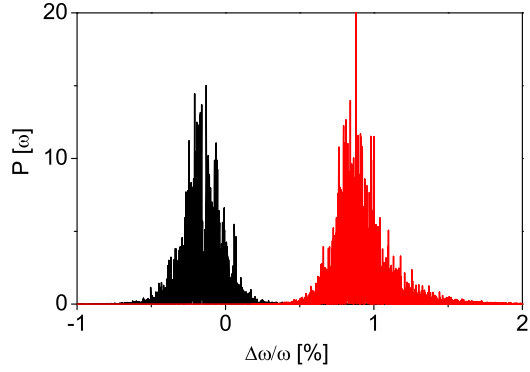}

\caption{
Normalized spectral power of the fundamental harmonic
for untapered (black) and tapered (red) case.
Output points correspond to the maximum
brilliance: $z = $ 53 m and $z = $ 80 m for untapered and tapered case,
respectively. Spectrum of untapered case is shifted by 1\% to the right-hand
side.
}
\label{fig:w1t-sat}
\end{figure}

\begin{figure}[tb]

\includegraphics[width=0.45\textwidth]{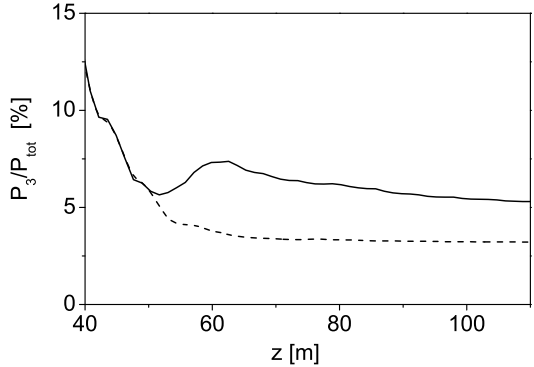}

\caption{
Contribution of the 3rd harmonic radiation power to the total radiation power
for untapered (solid curves) and optimized tapered case
(dashed curves).
}
\label{fig:p3toptot}
\end{figure}

\begin{figure}[tb]

\includegraphics[width=0.45\textwidth]{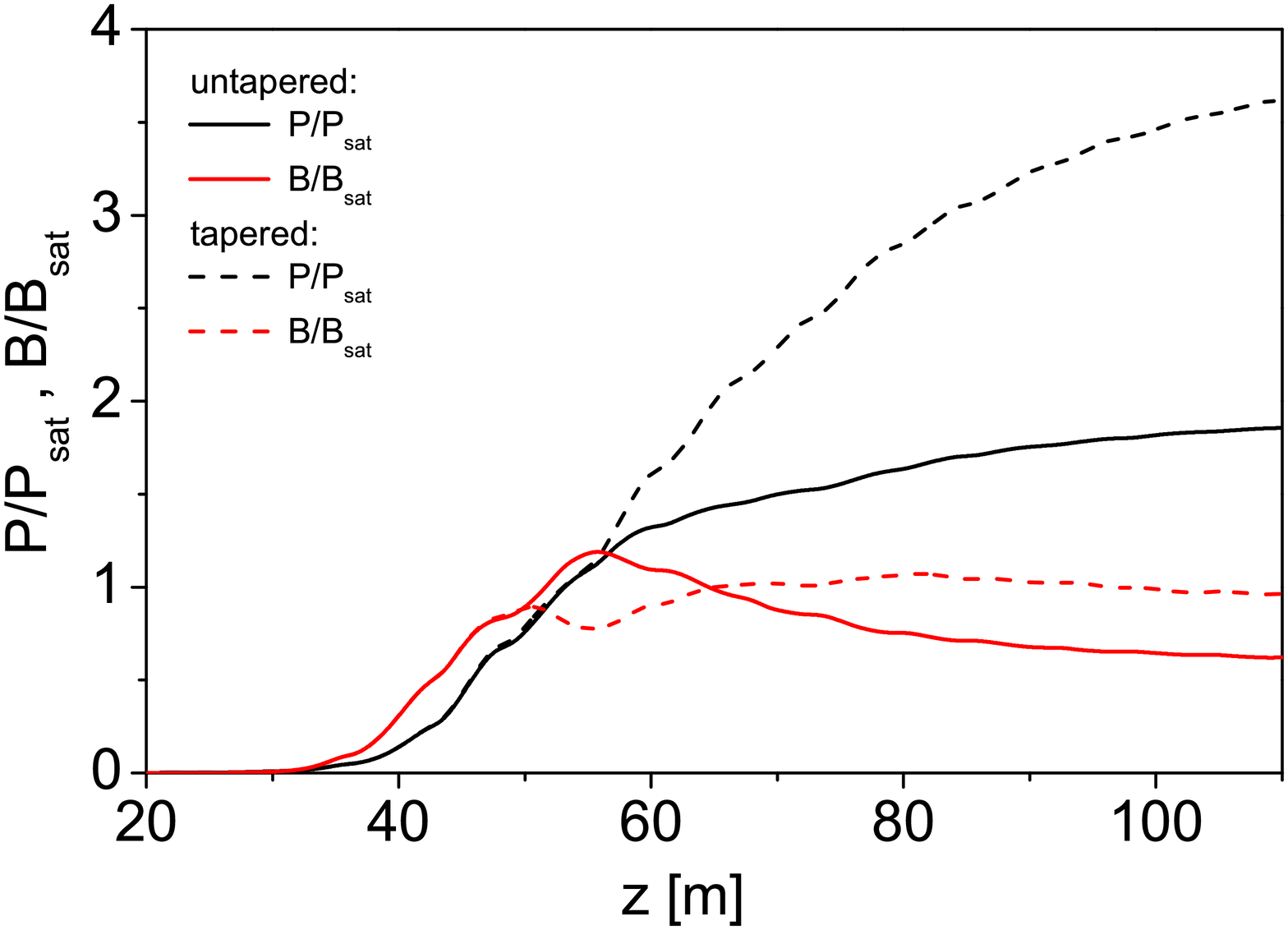}
\includegraphics[width=0.45\textwidth]{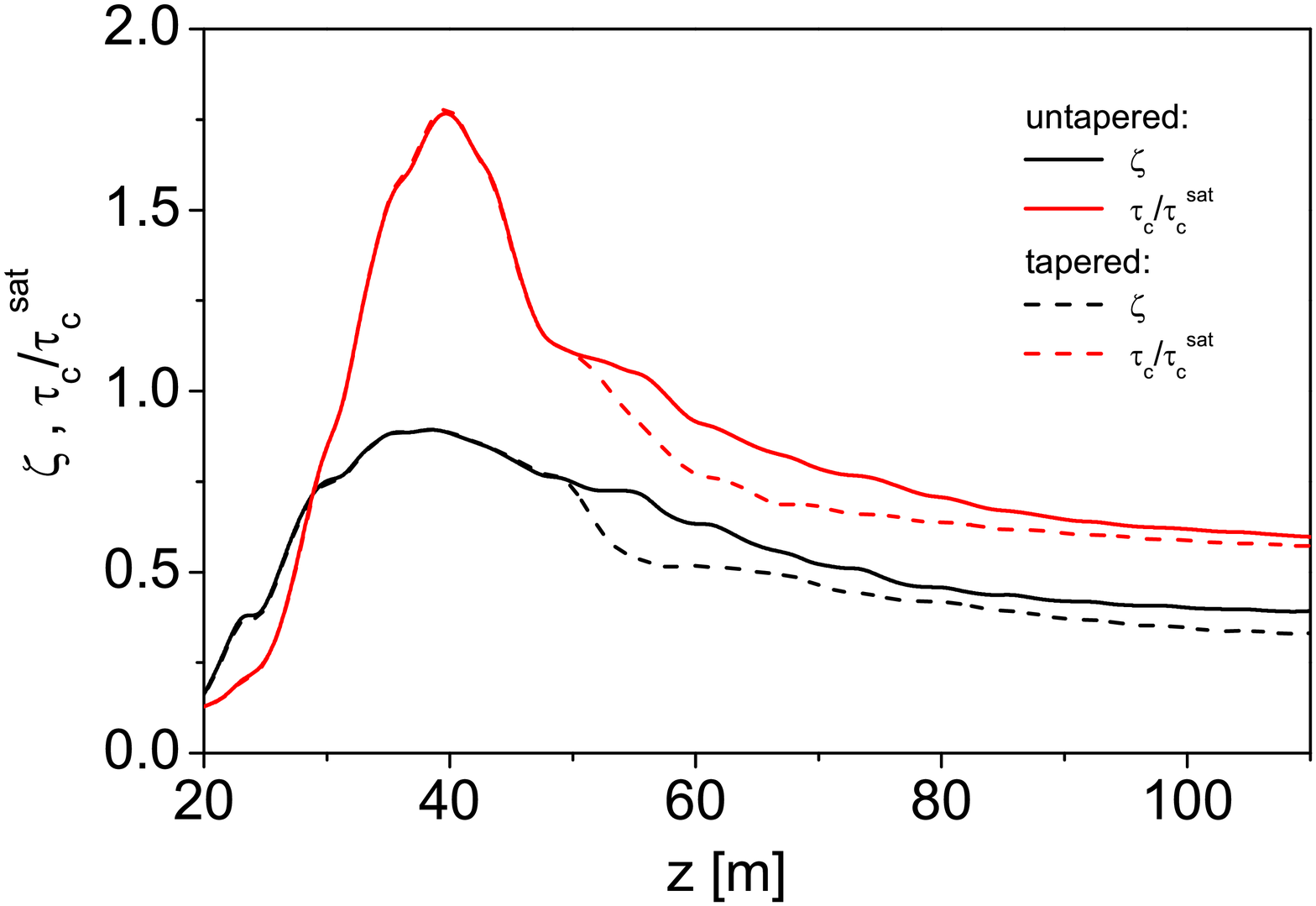}

\caption{
3rd harmonic:
evolution of the radiation power and brilliance (left plot) and of
coherence time and degree of transverse coherence (right plot) along the
undulator for untapered (solid curves) and optimized tapered case (dashed
curves). } \label{fig:pz3} \end{figure}

\begin{figure}[tb]

\includegraphics[width=0.45\textwidth]{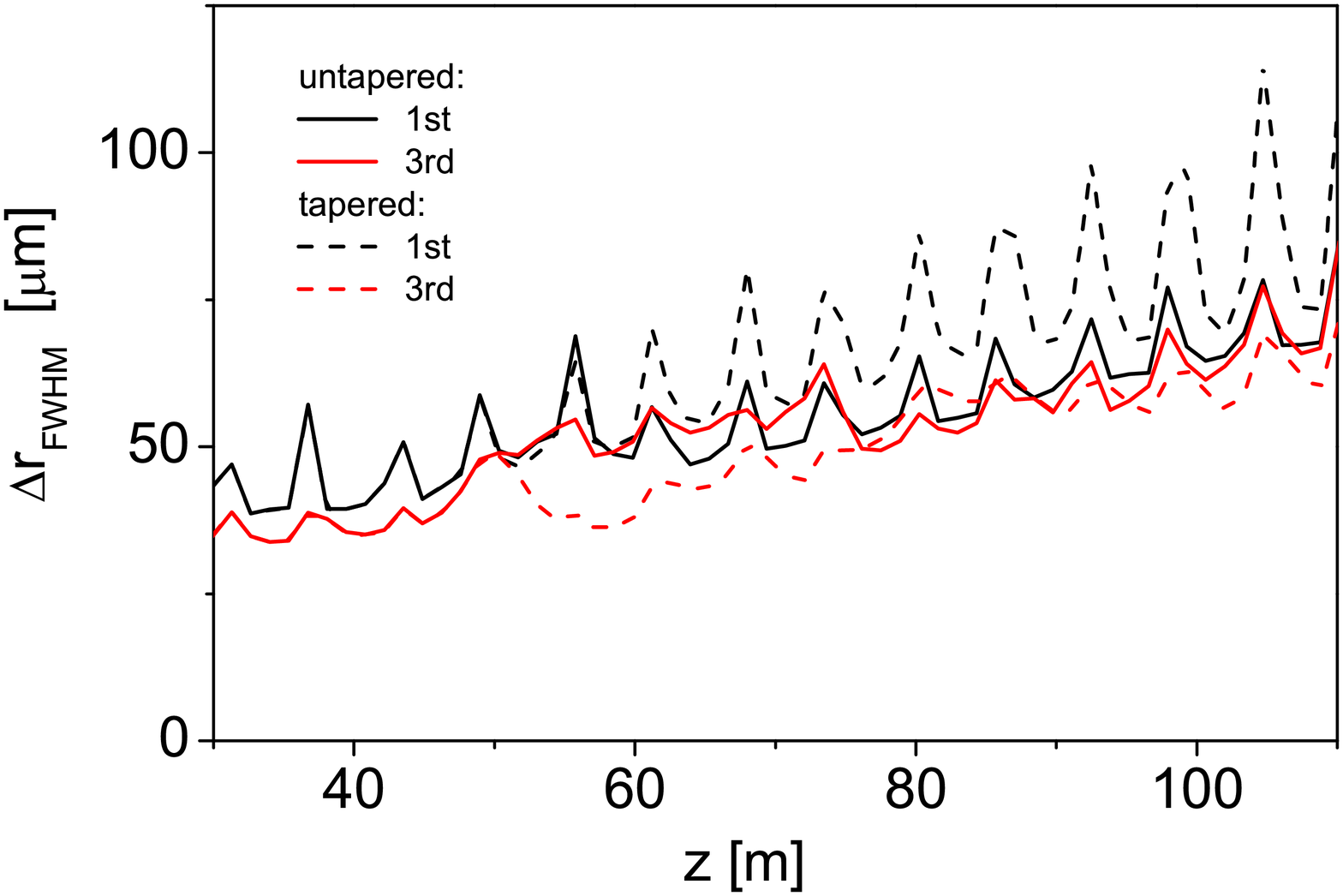}
\includegraphics[width=0.45\textwidth]{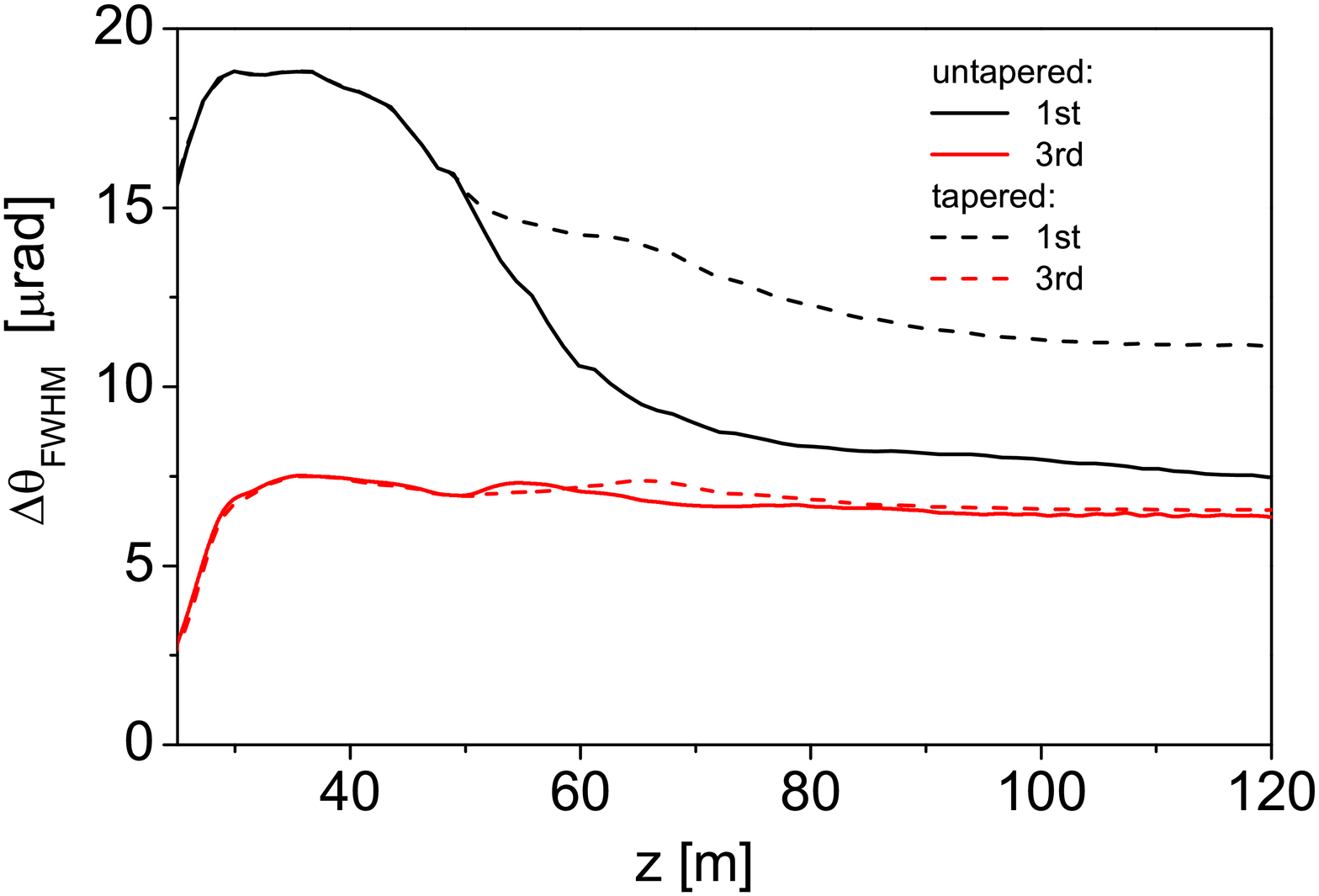}

\caption{
Evolution along the undulator of the FWHM spot size (left plot) and FWHM angular
divergence of the radiation in the far zone (right plot)
for untapered (solid curves) and optimized tapered case (dashed curves).
Black and red colors correspond to
the fundamental and the 3rd harmonic, respectively.
}
\label{fig:fwhm}
\end{figure}

\section{Results}

Simulations have been performed with the three-dimensional, time-dependent FEL
simulation code FAST \cite{coherence-oc,harm-stat-3d-fel2012,fast}. In our
simulation procedure we trace real number of electrons randomly distributed in
the six-dimensional phase space \cite{coherence-oc,harm-stat-3d-fel2012}. This
allows us to avoid any artificial effects arising from standard procedures of
macroparticle loading as we described earlier \cite{coherence-oc}. Simulations
of the FEL process have been performed for the case of a long bunch with the
uniform axial profile of the beam current. Output of the simulation code are
arrays containing complex values of the radiation field amplitudes. Then we
apply statistical analysis and calculate physical values as it has been defined
in the previous section. Such a model provides accurate predictions for the
coherence properties of the XFEL.

We perform comparative analysis of tapered and untapered case for the
parameters of the SASE3 undulator of the European XFEL. Undulator period is 6.8
cm, electron energy is 14 GeV, radiation wavelength is 1.55 nm. Undulator
consists of 21 modules, each is 5 meters long with 1.1 m long intersections
between modules. Parameters of the electron beam correspond to 0.25 nC case of
the baseline parameters of the electron beam: emittance 0.6 mm-mrad, rms energy
spread 2.5 MeV, peak beam current 5 kA \cite{desy11-152}. Average focusing beta
function is equal to 15 m. The value of the diffraction parameter is B = 1.1
which is close to the optimum conditions for reaching the maximum value of the
degree of transverse coherence \cite{coherence-oc}. Two cases were simulated:
untapered undulator, and the undulator optimized for maximum FEL efficiency
\cite{opt-tap-fel2014}.

Plots in Fig.~\ref{fig:pz1} show evolution along the undulator of the radiation
power, the degree of transverse coherence, the coherence time, and the
brilliance for the fundamental harmonic. For the case of untapered undulator
the coherence time and the degree of transverse coherence reach maximum values
in the end of the linear regime. Maximum brilliance of the radiation is
achieved in the very beginning of the nonlinear regime which is also referred
as the saturation point \cite{coherence-oc}. In the case under study the
saturation occurs at the undulator length of 53 m. Parameters of the radiation
at the saturation point are: the radiation power is 108 GW, the coherence time
is 1.2 fs, the degree of transverse coherence is 0.86, and the brilliance of
the radiation is equal to $3.8 \times 10^{22}$ photons/sec/mm$^2$/rad$^2$/0.1\%
bandwidth. The radiation characteristics plotted in Fig.~\ref{fig:pz1} are
normalized to the corresponding values at the saturation point.

General observations for the tapered regime are as follows. Radiation power
grows faster in than the untapered tapered case. The coherence time and the
degree of transverse coherence degrade, but a bit less intensive than in the
untapered case. Brilliance of the radiation for the tapered case saturates at
the undulator length of 80 m, and then drops down gradually. Benefit of the
tapered case against untapered case in terms of the radiation brilliance is
factor of 3, and it is mainly defined by the corresponding increase of the
radiation power. Coherence properties of the radiation in the point of the
maximum brilliance of the tapered case are worse than those of the untapered
SASE FEL in the saturation point: 0.86 to 0.68 for the degree of transverse
coherence, and 1 to 0.86 in terms of the coherence time. Figure
\ref{fig:w1t-sat} presents normalized spectral power of the fundamental
harmonic for untapered (black) and tapered (red) case. Output points correspond
to the maximum brilliance: $z = $ 53 m and $z = $ 80 m for untapered and
tapered case, respectively. Spectrum of the untapered case is
more narrow and does not contain spanning tails.

We already mentioned that parameters of SASE3 FEL are such that we expect
significant increase of the radiation power in the higher odd harmonics due to
the mechanism of the linear harmonic generation. This happens due to small value
of the diffraction parameter and high quality of the electron beam (small
emittance and energy spread) \cite{sy-harm}. The results of numerical
simulations presented in Fig.~\ref{fig:p3toptot} shows that for the untapered
case the contribution of the 3rd harmonic to the total power is about 6\% in
the saturation point. Note that the mechanism of the nonlinear harmonic
generation results in the value of 2\% only. High value of the 3rd harmonic
radiation power should be the subject of concern for the planned user
experiments. It can constitute harmful background, or it can be used in
the pump-probe experiments. Mechanisms to control the 3rd harmonic radiation are
now the subject of dedicated studies \cite{harm-control-2014}.

The evolution along the undulator of the radiation
power, the degree of transverse coherence, the coherence time, and the
brilliance for the 3rd harmonic is shown in Fig.~\ref{fig:pz3}. The values are
normalized to the corresponding values of the 3rd harmonic at the saturation
point of the fundamental harmonic at the undulator length of 53 m. Normalizing
factors are 6.6 GW for the radiation power, 05 fs for the coherence
time, and 0.72 for the degree of transverse coherence. There is
interesting observation that the brilliance of the radiation of the 3rd harmonic
does not differ significantly for the tapered and untapered cases. In the case
of the optimized tapered undulator the relative contribution of the 3rd harmonic
to the total power is visibly less while the absolute power is higher than for
the untapered case. Coherence properties of the 3rd harmonic for untapered case
are a bit better than those for the tapered case.

To make our paper complete, we conclude with presenting in Fig.~\ref{fig:fwhm}
of the evolution along the undulator of the FWHM spot size and the FWHM angular
divergence of the radiation in the far zone. The cone of the fundamental
harmonic radiation in the far zone is visibly wider for the tapered case. Also,
phase front of the radiation is quite different for the tapered and untapered
case. This is a hint for careful design of the optical transport system capable
effectively handle both, untapered and tapered options.

\section{Summary}

Application of the undulator tapering has evident benefit for SASE3 FEL operating
in the wavelength range around 1.6 nm. It is about factor of 6 in the pulse
radiation energy with respect to the saturation regime, and factor of 3 with
respect to the radiation power at a full length. General feature of tapered
regime is that both, spatial and temporal coherence degrade in the nonlinear
regime, but more slowly than for untapered case. Peak brilliance is reached in
the middle of tapered section, and exceeds by a factor of 3 the value of the
peak brilliance in the saturation regime. The degree of transverse coherence at
the saturation for untapered case is 0.86. Degree of transverse coherence for
the maximum brilliance of the tapered case is 0.66. Coherence time falls by
15\%. At the exit of the undulator the degree of transverse coherence for
the tapered case is 0.6, and coherence time falls by 20\%. Radiation of the 3rd
harmonic for both, untapered and tapered cases, exhibit
nearly constant brilliance and nearly constant contribution to the total power.
Coherence time of the 3rd harmonic for the tapered case approximately scales
inversely proportional to harmonic number, as in untapered case.

\end{document}